\documentclass{ws-ijmpa}
\usepackage{bm}
\usepackage{graphicx}

\begin{document}

\markboth{M.~Wingate et al.}
{$B$ Decays on the Lattice and Results for Flavor Phenomenology}

\catchline{}{}{}{}{}
%
%%%%%%%%%%%%%%%%%%%%%%%%%%%%%%%%%%%%%%%%%%%%%%%%%%%%%%%%%%%%%%%%%%%%

\title{B DECAYS ON THE LATTICE AND RESULTS FOR PHENOMENOLOGY}

\author{MATTHEW WINGATE}
\address{Institute for Nuclear Theory, University of Washington\\
Seattle, WA, 98195-1550, USA}

\author{CHRISTINE DAVIES}
\address{Department of Physics \& Astronomy, University of Glasgow\\
Glasgow, G12 8QQ, UK}

\author{ALAN GRAY, EMEL GULEZ, JUNKO SHIGEMITSU}
\address{Department of Physics, Ohio State University\\
Columbus, OH 43210, USA}

\author{G.\ PETER LEPAGE}
\address{Laboratory for Elementary Particle Physics, Cornell University\\
Ithaca, NY 14853, USA}

\maketitle

\begin{abstract}
Lattice Monte Carlo simulations now include the effects
of 2 light sea quarks and 1 strange sea quark through the
use of an improved staggered fermion action.  Consequently,
results important to phenomenology are free of the approximate
10\% errors inherent in the quenched approximation.  
This talk reports on calculations of the $B$ and $B_s$ decay constants
 and $B \to \pi \ell \nu$ form factors.  Accurate determinations
of these quantities will lead to tighter constraints on CKM matrix
elements.

\end{abstract}

%----------------------------------------------------------------o

\section{Introduction}

Lattice QCD results
for hadronic matrix elements governing neutral $K$ and $B$ mixing
are important elements in the quest to determine the CKM parameters
$\overline{\rho}$ and $\overline{\eta}$.\cite{Lubicz:2004nn}
{\it Ab initio} calculations of form factors for $K$, $D$, and $B$ 
exclusive semileptonic decays can be combined with experiment
to obtain CKM matrix elements, complementing other determinations,
e.g.\ through inclusive decays.  Furthermore, many of the lattice
calculations in the charm sector will be testable at current
experiments, notably CLEO-c.\cite{Shipsey:2004wz}

The effects of light sea quark masses are now included in present
state-of-the-art lattice simulations.  The first round of results
showed that cleanly computable, or ``golden,''
quantities agree with experiment within the few percent lattice
uncertainties.\cite{Davies:2003ik}  
This is in sharp contrast to quenched lattice calculations,
which neglect sea quark effects and have 10-20\% discrepancies. 
Having shed the quenched approximation and its errors, 
lattice calculations are
in a position to make an even greater impact on flavor phenomenology.

This talk briefly summarizes our recent work on the $B$ and $B_s$
decay constants and the form factors describing $B\to\pi\ell\nu$ decay.
%A report on the current status of the field was delivered at 
%{\it Lattice 2004}.\cite{Wingate:2004xa}

\section{Decay Constants}

The $B_s$ decay constant is straightforward to compute since no
extrapolations in the valence quark masses are needed.
We recently computed it to be\cite{Wingate:2003gm}
\begin{equation}
f_{B_s} ~=~ 260 \pm 7 ~\mathrm{(stat)}~ \pm 28~\mathrm{(sys)} 
~\mathrm{MeV} \, .
\end{equation}
This represents the current state-of-the-art because 
effects of 2 light and 1 strange sea quarks are included.
The dominant systematic uncertainty is estimated to come from
higher order terms in the perturbative matching of lattice NRQCD
to continuum QCD.\cite{Gulez:2003uf}  
Prospects for reducing that uncertainty lie
in a future 2-loop matching calculation or a partially nonperturbative
estimate.

The calculation of $f_B$ requires significantly more work.  
The light quark mass extrapolation is done using
chiral perturbation theory; however, the simulations must be
performed at light enough masses that the leading order terms
are sufficient.  
Fig.~1 (left) shows our progress computing $f_B$. 
We vary both the valence and sea quark masses which will permit
a partially quenched analysis.\cite{Wingate:2003ni}  Recent
work has focused on techniques to improve the signal-to-noise ratio
for the correlation functions we compute.\cite{Gray:2004hd}
The leftmost 2 points in Fig.~1 (left) illustrate a dramatic
improvement in statistical resolution.  The large uncertainty
due to perturbative matching will cancel in the ratio $f_{B_s}/f_B$,
allowing us to tighten the $|V_{td}|$ constraint.

\begin{figure}
\centerline{\hbox{
\includegraphics[width=6cm,origin=c,angle=270]{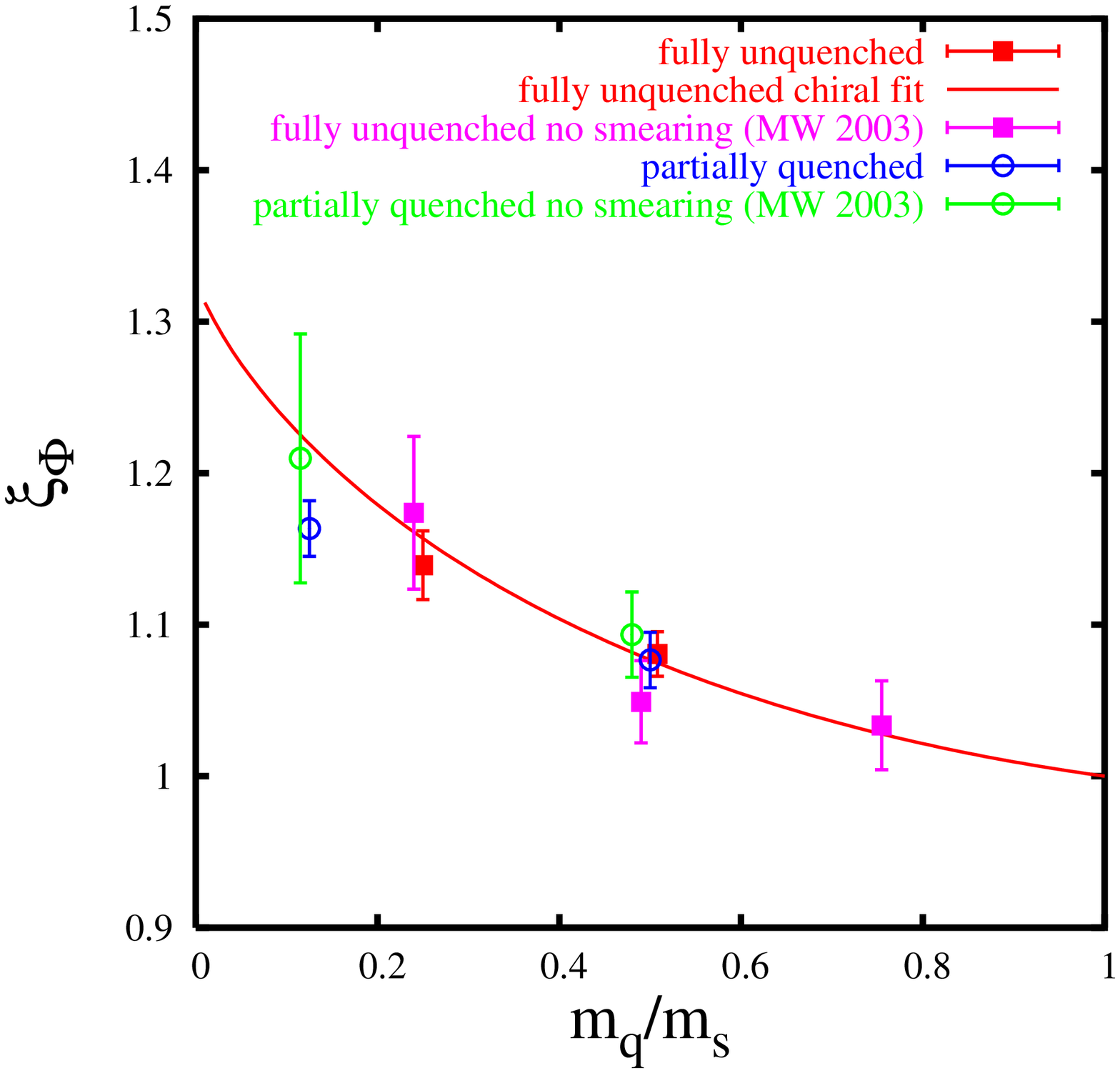}
~~\includegraphics[width=6cm]{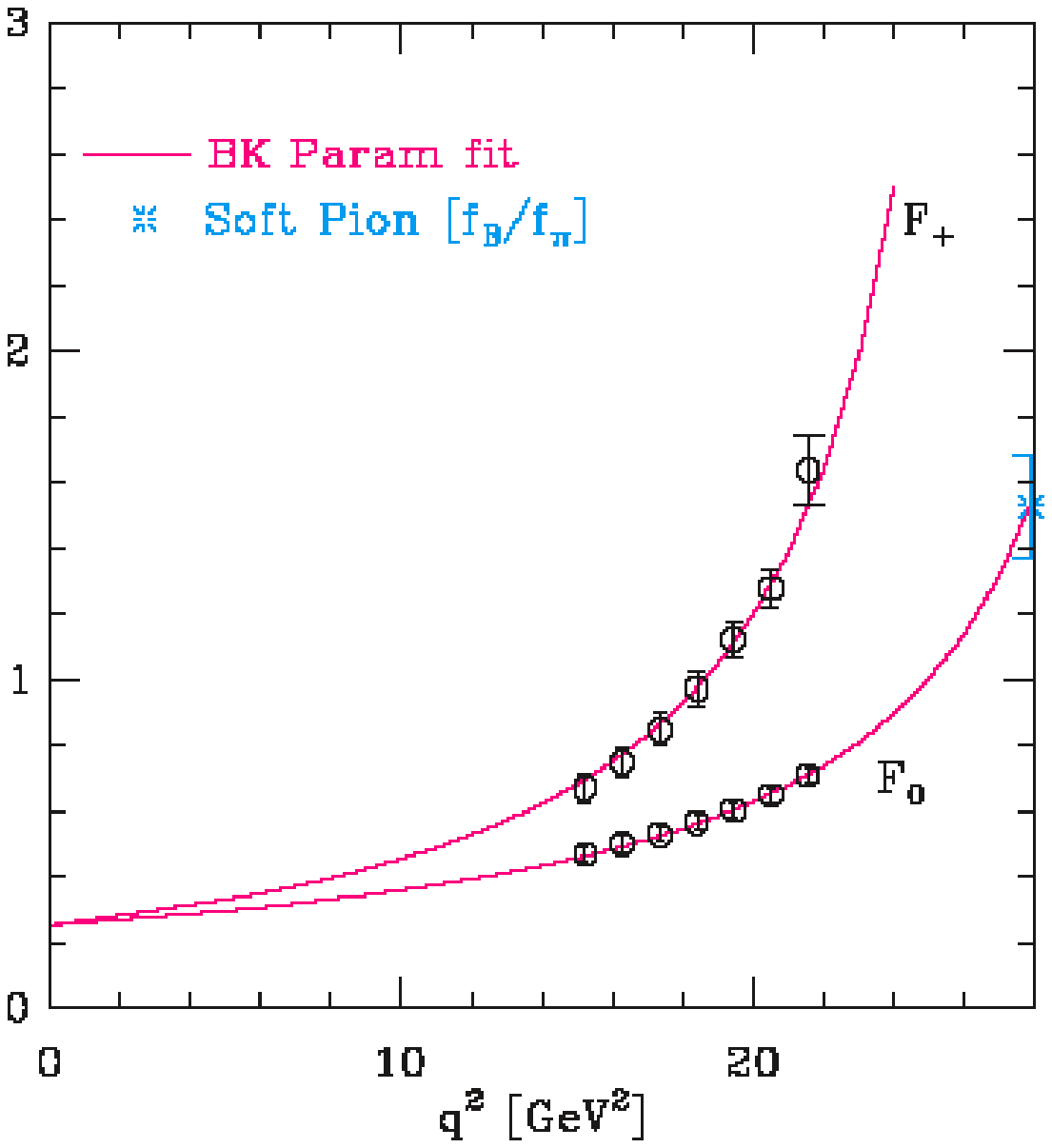}
}}
\vspace*{8pt}
\caption{\label{fig1} Left: Progress computing the ratio of decay 
constants $\xi_\phi \equiv (f_{B_s}\sqrt{m_{B_s}})/(f_B\sqrt{m_B})$; 
the solid curve illustrates unquenched chiral behavior.
Right: Preliminary results for $B\to \pi\ell\nu$ form factors;
the data are fit well by the Be\'cirevi\'c - Kaidalov ansatz.}
\end{figure}

\section{Semileptonic Form Factors}

We recently presented preliminary results for the $B\to\pi\ell\nu$
form factors.\cite{Shigemitsu:2004ft}  Fig.~1 (right) shows the
form factors $f_+$ and $f_0$ as functions of momentum transfer.
Simulations can presently be carried out only at large $q^2$.  
As $q^2$ decreases, the pion momentum increases and discretization
errors which scale like $(p_\pi a)$ become sizable. 
After extrapolating the light quark mass to its
physical value, the form factors are fit well by the Be\'cirevi\'c
-  Kaidalov ansatz.\cite{Becirevic:1999kt}  
One can integrate the form factors and combine
the result with the $B^0$ lifetime and the experimental branching ratio 
for $B^0 \to \pi^- \ell^+ \nu$ to obtain $|V_{ub}|$.\cite{Athar:2003yg}
  The lattice errors
can be decreased if the integration in done only over a range of $q^2$
for which we have data; this result is then combined with the
branching ratio for decays with $q^2 \ge 16$ GeV${}^2$.  Unfortunately
applying the momentum cut increases the experimental uncertainty.
The preliminary results for both procedures are
\begin{equation}
|V_{ub}| ~=~ \left\{ \begin{array}{lcl}
3.86\pm 0.32 \;\mathrm{(expt)}\pm 0.58 \;\mathrm{(latt)} \times 10^{-3} 
& \hspace{3mm} & 
0 \le q^2 \le q^2_\mathrm{max} \nonumber \\
3.52 \pm 0.73\;\mathrm{(expt)} \pm  0.44\;\mathrm{(latt)} \times 10^{-3} & &
16~\mathrm{GeV}^2 \le q^2 \le q^2_\mathrm{max}
\end{array} \right. \, .
\end{equation}

%----------------------------------------------------------------o
\section{Summary}

Unquenched lattice calculations with quarks light enough to 
 overlap with the region
of validity of chiral perturbation theory will lead to much more
accurate results for hadronic matrix elements.  The first round
of results are already promising.  Here we presented results
for $f_{B_s}$ and $B\to \pi \ell\nu$ form factors.  
In the absense of quenched artifacts, we are able to study better the
leading systematic uncertainties and find ways to reduce them.
Ongoing efforts will lead to more accurate determinations of $f_B$ and the 
mixing parameters $\mathrm{B}_B$ and $\mathrm{B}_{B_s}$.

%----------------------------------------------------------------o
%\section*{Acknowledgments}

This work was supported in part by DOE 
(DE-FG02-00ER41132), NSF, and PPARC.  Simulations were performed
at NERSC.

%----------------------------------------------------------------o

\end{document}